\documentclass[%
 reprint,
 amsmath,amssymb,
 aps,
]{revtex4-2}

\usepackage{textgreek}
\usepackage{graphicx}
\usepackage{dcolumn}
\usepackage{bm}

\begin{document}

\preprint{APS/123-QED}

\title{Supercurrent Noise in a Phase-Biased Superconductor-Normal Ring in Thermal Equilibrium}

\author{Ziwei Dou\textsuperscript{1}, Xavier Ballu\textsuperscript{1}, Quan Dong\textsuperscript{2}, Yong Jin\textsuperscript{3}, Richard Deblock\textsuperscript{1}, Sandrine Autier-Laurent\textsuperscript{1}, Sophie Gu\'{e}ron\textsuperscript{1},  
\\H\'{e}l\`{e}ne Bouchiat\textsuperscript{1} , and Meydi Ferrier\textsuperscript{1}}
\affiliation{1. Universit\'e Paris-Saclay, CNRS, Laboratoire de Physique des Solides, 91405, Orsay, France
\\2. CryoHEMT Electronics for Scientific Research, 91400, Orsay, France
\\3. Université Paris-Saclay, CNRS, Centre de Nanosciences et de Nanotechnologies, 91120, Palaiseau, France}




\date{\today}

\begin{abstract}
In superconductor-normal-superconductor (SNS) junctions, dissipationless supercurrent is mediated via Andreev bound states (ABSs) controlled by the phase difference between the two superconductors. Theory has long predicted significant fluctuations, and thus a noise, of such supercurrent in equilibrium, due to the thermal excitation between the ABSs. Via the fluctuation-dissipation theorem (FDT), this leads paradoxically to a dissipative conductance even in the zero frequency limit. In this article, we directly measure the supercurrent noise in a phase-biased SNS ring inductively coupled to a superconducting resonator. Using the same setup, we also measure its admittance 
and quantitatively demonstrate the FDT for any phase. The dissipative conductance shows an 1/T temperature dependence, in contrast to the Drude conductance of unproximitized metal. This is true even at phase $\pi$ where the Andreev spectrum is gapless. Supported by linear response theory, we attribute this to the enhanced current correlation between the ABSs symmetrical across the Fermi level. Our results provide insight into the origin of noise in hydrid superconducting devices which may be useful for probing topologically protected ABSs.     
\end{abstract}


\maketitle



\noindent To what extent is a dissipationless system noisy? This paradoxical question is of fundamental importance for the ongoing research which combines superconducting and topological materials with dissipationless features. A paradigmatic example is a Josephson junction, where two superconducting electrodes are coupled by a tunnel barrier. A well-known property of such a junction is the dissipationless supercurrent that flows with no voltage bias. As a consequence, no fluctuations can be detected in the supercurrent at frequency or temperature much smaller than the superconducting gap $\Delta$ \cite{ROGOVIN_fluc}. This is intuitively consistent with the fluctuation-dissipation theorem \cite{Callen-Welton_noise, Kubo_fluc_diss_theorem} (FDT) stating that dissipation and current noise are proportional in equilibrium. In superconductor-normal-superconductor (SNS) junctions, the tunnel barrier is replaced by a normal metal. There, a supercurrent is carried by the phase-coherent Andreev bound states (ABSs), whose spectrum varies with the phase difference $\varphi$ between the two superconducting electrodes. In a long diffusive junction, the Andreev spectrum displays an induced ``minigap" $E_g \ll \Delta$ which is maximal at phase 0 and closes at phase $\pi$ \cite{minigap_theory,minigap_STM}. Greater noise has been predicted in such junctions since there exists a regime where $E_g < k_BT \ll \Delta$ and transitions of the ABSs across the minigap occur more frequently \cite{averin_Imam, spivakdissipation, rodero_thermal_noise}. 

\noindent The supercurrent fluctuations are characterized by their noise power spectrum $S_I(f)$, which is the Fourier transform of the time correlation of current. From the FDT, a dissipation is expected: it takes the form of a conductance $G$ and obeys the relation $S_{I}=4k_BTG$, where $T$ is the temperature. The theoretical prediction of finite fluctuations down to zero frequency \cite{averin_Imam, spivakdissipation, rodero_thermal_noise} thus entails a non-zero $G$, in apparent contradiction with the dissipationless character of the supercurrent. The resolution of this paradox calls for a rigorous definition of the conductance. Naively, $G$ is the ratio of the current to voltage in the linear regime and requires a voltage biasing scheme. For a low impedance system such as an SNS junction, such a biasing scheme is challenging, particularly because even an infinitely small dc voltage bias drives the junction out of equilibrium due to the ac Josephson effect \cite{spivakdissipation}. A more suitable approach is to adopt a phase-biasing scheme using an SNS ring. There the phase can be controlled by the dc magnetic flux via $\varphi = 2\pi\Phi/\Phi_0$ ($\Phi_0= h/2e$). Meanwhile, the conductance can be accessed by adding a small ac flux $\delta\Phi \ll \Phi_0$ at frequency $f$, which is produced for example by a superconducting resonator inductively coupled to the ring. The magnetic susceptibility $\chi = \delta i/\delta\Phi$ can thus be measured. Generally, $\chi$ is complex due to the finite relaxation time of the ABSs in response to microwave or thermal excitation. It thus results in an admittance $Y = -i\chi/2\pi f$ whose conductance is the 
real part $G=Re(Y)$ \cite{Pauli_ac_chi,tikhonov_admittance,bastien_prb}.

\noindent In this work, we have directly measured the long predicted supercurrent noise of a phase-biased SNS ring. We have also measured its conductance, and quantitatively verified the FDT over the full range of phase between 0 and $2\pi$. Moreover, the experiment identifies an $1/T$ temperature dependence of the conductance in contrast to the constant Drude conductance. Surprisingly, it is true even at $\varphi = \pi$ when the minigap closes. Supported by linear response theory, we show that such behavior is a manifestation of the enhanced current correlation between the ABSs symmetrical across the Fermi level.

\section*{Measurement method}

\begin{figure}[]
  \centering
    \includegraphics[width= 0.45\textwidth]{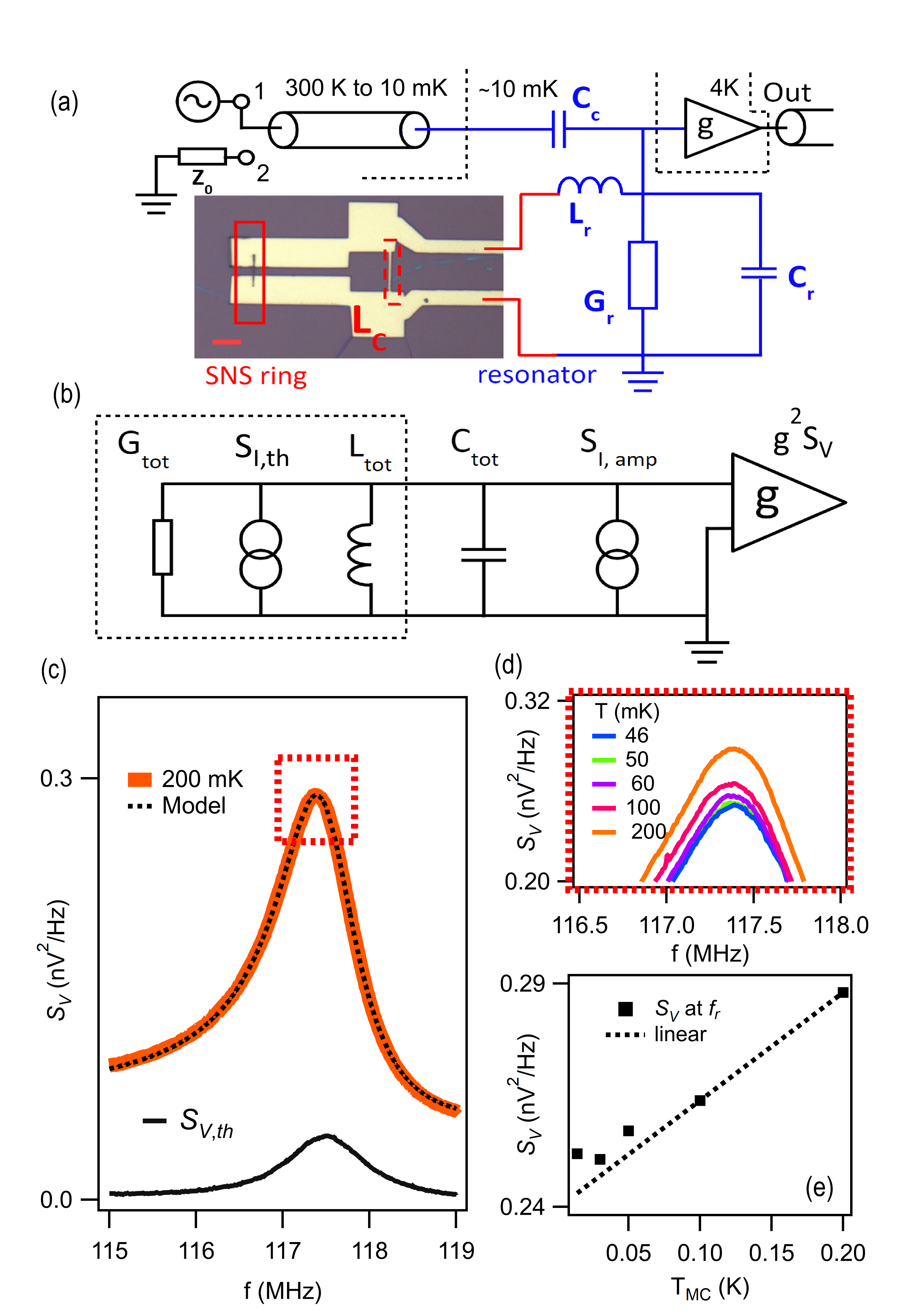}
\caption{Noise spectrum of the measurement system: (a) Schematic of the measurement setup. Noise or transmission measurement is performed using the same setup by connecting the input to a grounded $Z_0$ (port 2) at 300 K or to an rf source (port 1), respectively. The superconducting resonator (blue) is inductively coupled to the SNS ring (optical image, scale bar: 5 \textmu m). Solid rectangle: the normal metal wire. Dashed rectangle: the coupling inductance $L_c$. The resonator (at $\sim$ 10 mK) is directly connected to a low-temperature amplifier at 4 K with the gain $g$. (b) Effective circuit modeling of (a). The total device (the resonator and the SNS ring, dashed box) is modeled as a parallel-GLC circuit. See text. (c) Measured voltage noise spectrum $S_V(f)$ at $T$ = 200 mK (red), fitted to Eq. \ref{eq:Sv-f} (black dashed). The thermal contribution $S_{V,th}$ is plotted as the black continuous line. (d) $S_V(f)$ at $\varphi$ =  0 around $f_{r}$, showing temperature dependence. (e) $S_V(f_{r})$ extracted from (d) versus the mixing chamber temperature $T_{MC}$. The linear relation (dashed line) is extrapolated using the high temperature data.}
\label{fig:sys-charact}
\end{figure}

\noindent Fig. \ref{fig:sys-charact}(a) displays the schematic of the measurement setup, which can be configured to measure both the noise spectrum and the transmission coefficient. For transmission measurement, the input port is connected to an rf source with sufficiently small power and the transmitted signal at the output port is down-converted to obtain the transmission coefficient $\Gamma(f)$. For noise measurement, the input port is simply grounded via $Z_0$ = 50 {\textOmega} and the voltage noise spectrum $g^2S_{V}(f)$ at the output is recorded, where $g$ is the gain of the amplifier. Each spectrum is averaged and filtered according to the procedure detailed in \cite{supp}. The optical image shows the SNS ring, formed by the normal metal weak link on the left (solid rectangle) enclosed by the coupling inductance $L_c$ on the right (dashed rectangle). The dc magnetic flux $\Phi$ through the loop is used to phase-bias the junction via $\varphi = 2\pi\Phi/\Phi_0$. The normal wire is made by evaporating a Ti/Au of 5nm/100nm thickness on an undoped silicon substrate, with a length $L =$ 1.5 \textmu m and a width $W =$ 100 nm. The rest of the ring is formed by 80 nm thick superconductor molybdenum-rhenium (MoRe). The SNS ring is probed via the superconducting resonator. The resonator is coupled to the input transmission line by a coupling capacitance $C_c$, and is directly connected to a home-made HEMT amplifier at 4 K \cite{YongJin_hemt}. The small $C_c$ ($\sim$ 1 pF) and the large amplifier impedance ($>$ 1 G{\textOmega}) are chosen to preserve the resonator's quality factor. To reduce the input capacitance caused by the Miller effect, $g$ is limited around 1 \cite{pozar2011microwave}. Therefore, instead of amplification, the amplifier is used to match the impedance between the resonator and the external circuit. 

\noindent From Fig. \ref{fig:sys-charact}(b), the paralleled $G_{tot}$, $L_{tot}$ and $C_{tot}$ models the total system of the SNS ring plus the resonator. $S_{I,th}$ is the total current noise originated from the system. We note that, in the parallel GLC model, larger $G$ corresponds to greater dissipation. The contributions from the resonator and the ring can be separated \cite{supp}:

\begin{align}
\begin{split}
   &\frac{1}{L_{tot}}(\Phi) =  \frac{1}{L_{reso}} + \kappa \frac{1}{L_{ring}}(\Phi)\\
   &G_{tot}(\Phi) = G_{reso} + \kappa G_{ring}(\Phi)\\
   &S_{I,th}(\Phi) = 4k_BTG_{reso} + \kappa S_{I,ring}(\Phi)\\
\end{split}
\label{eq:GtotLtot} 
\end{align}

\noindent where the resonator parameters $1/L_{reso}$ and $G_{reso}$ are constant while $1/L_{ring}$ and $G_{ring}$ are phase-dependent. $\kappa = (L_c/L_{reso})^2$ is the inductive coupling coefficient and is a small factor in the order of 10\textsuperscript{-5}. For the resonator, we can directly write its thermal noise as $4k_BTG_{reso}$ \cite{resoFDT}. It is important that we do not assume such relation between $S_{I, ring}(\Phi)$ and $G_{ring}(\Phi)$. Indeed, the central goal of this article is to demonstrate that these two quantities, both due to the ABS dynamics, are also linked by the FDT.

\noindent As will be shown later, $1/L_{ring}(\Phi)$ and $G_{ring}(\Phi)$ are obtained by the phase-dependent resonance frequency and quality factor of $\Gamma(f)$ \cite{bastien_prb}. $S_{I,th}$, on the other hand, is extracted from the voltage noise $g^2S_V(f)$ measured at the amplifier output. From Fig. \ref{fig:sys-charact}(b), the relation between $S_{I,th}$ and $S_V(f)$ is \cite{supp}:

\begin{align}
\begin{split}
   &S_{V}(f) = S_{V,th}(f)+S_{V,amp}(f)  \\
   &S_{V,th} = \frac{S_{I,th}}{|Y_{tot}|^2}, S_{V,amp} = \frac{S_{I,amp}(Y_{tot})}{|Y_{tot}|^2} \\
\end{split}
\label{eq:Sv-f} 
\end{align}

\begin{figure*}[]
  \centering
    \includegraphics[width= 1\textwidth]{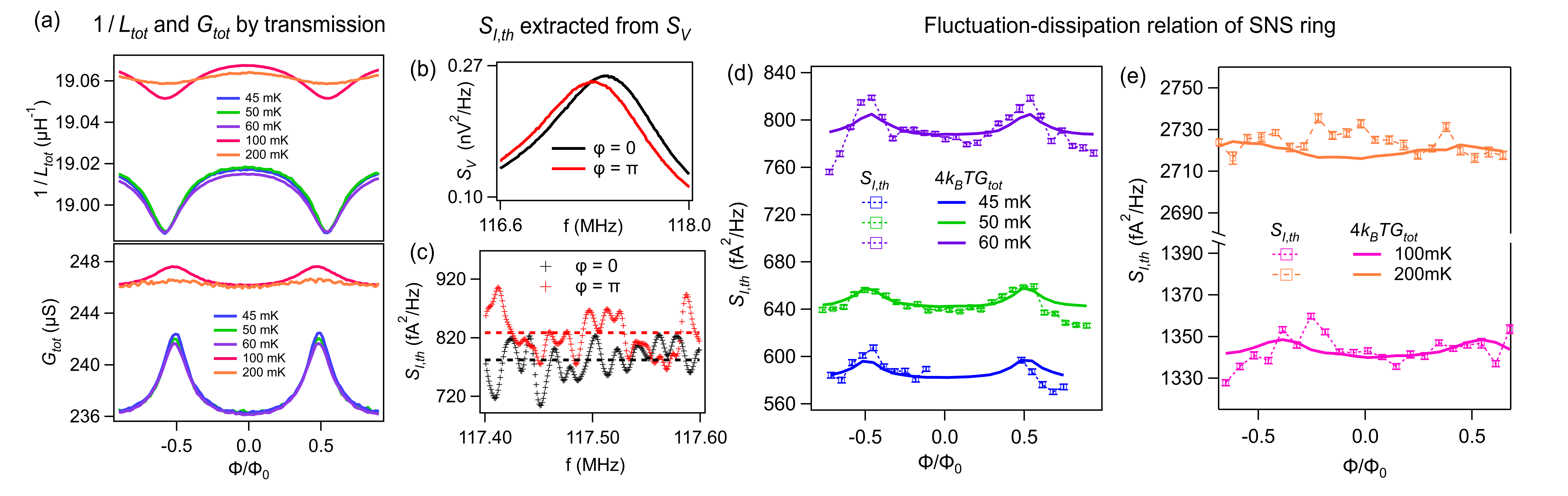}
\caption{Phase-dependent supercurrent noise and demonstration of FDT for the SNS ring: (a) $1/L_{tot}$ and $G_{tot}$ measured by transmission for all temperatures. (b) Measured $S_V(f)$ at $\varphi$ = 0 (black) and $\pi$ (red). (c) Extracted $S_{I,th}(f)$ at $\varphi$ = 0 (black) and $\pi$ (red), with the mean value marked by the dashed lines. (d,e) Fluctuation-dissipation relation of the total device, comparing the mean values of the extracted noise $S_{I,th}$ (squares) and $4k_BTG_{tot}$ (solid lines) using the transmission data in (a). The phase-independent background $4k_BTG_{reso}$ is the thermal noise from the resonator. The phase-dependent part $\kappa S_{I,ring}$ is due only to the supercurrent noise of the SNS ring.}
\label{fig:invLr-Gr_compare}
\end{figure*}

\noindent where $Y_{tot} = G_{tot}+1/(j2\pi f L_{tot}) + j2\pi f C_{tot}$ is the total admittance of the device. The noise of the amplifier is modeled as $S_{I,amp}$, which depends on both $Y_{tot}$ and the phase-independent internal parameters \cite{supp}. From Eqs. \ref{eq:GtotLtot} and \ref{eq:Sv-f}, to accurately extract $S_{I,th}$, both the resonator parameters $1/L_{reso}$ and $G_{reso}$ and the amplifier noise $S_{I,amp}$ should be calibrated. Since we are predominantly interested in the phase dependent part of the noise, we re-calibrate those parameters at phase 0 for each temperature which is then kept constant for the other phases. This is done by assuming the FDT for the phase-independent elements of the setup $S_{I,th}(0)=4k_BTG_{th}(0)$ and fitting the measured $S_V(f)$ to Eqs. \ref{eq:Sv-f} \cite{supp}. Such calibration scheme is required to compensate any drift of the amplifier and resonator parameters with temperature or even time, which is indeed seen later in Fig. \ref{fig:invLr-Gr_compare}(a). Nevertheless, we check that the maximum error made on $G_{reso}$ in the order of $\kappa G_{ring}$ does not affect the validity of our phase-dependent results \cite{supp}. 

\noindent  The superb quality of the fit obtained in Fig. \ref{fig:sys-charact}(c) confirms the relevance of our model. By subtracting the amplifier contribution $S_{V,amp}$, the thermal noise $S_{V,th}$ contributes to roughly 20\% of $S_V$. We also see clearly the temperature dependence of the noise on resonance in Figs  \ref{fig:sys-charact}(d) and (e). Assuming a linear dependence at the highest $T$ [dashed line Fig. \ref{fig:sys-charact}(e)], the data below 50 mK digress slightly from the such relation, indicating a different electronic temperature from that of the mixing chamber. Using the extrapolated linear relation, the electronic temperatures are calibrated in \cite{supp} whose values are noted in Fig. \ref{fig:sys-charact}(d) and the later figures. 

\section*{Experimental Validation of FDT}

\noindent The main result of this work is the experimental demonstration of the FDT for an SNS ring. It consists of measuring independently $S_{I,ring}$ from the voltage noise spectrum and $G_{ring}$ from the transmission coefficient and check the relation $S_{I,ring}(\Phi)=4k_BTG_{ring}(\Phi)$ at all phases. First we measure the admittance of the SNS ring via transmission coefficient according to the following relations \cite{supp,bastien_prb}: 

\begin{align}
\begin{split}
    &\frac{\kappa}{L_{ring}}(\Phi) = \frac{2}{L_{reso}}\frac{\delta f_r(\Phi)}{f_r} \\
   &\kappa G_{ring}(\Phi) = \frac{1}{2\pi f_rL_{reso}}\delta \left(\frac{1}{Q}\right)(\Phi) \\
\end{split}
\label{eq:LG_fQ} 
\end{align}

\noindent where $f_r$ and $Q$ are the resonance frequency and the quality factor respectively. Following the method described in \cite{supp, bastien_prb}, as $\Phi$ is swept, we simultaneously record $\delta f_r$ and $\delta (1/Q)$ using a phase-locked feedback loop which maintains the resonator on resonance \cite{bastien_prb}. Since the above method only determine the phase-dependence of the ring admittance, we can set $\kappa/L_{ring}(0)$ and $\kappa G_{ring}(0)$ as 0. Adding to $1/L_{tot}(0)$ and $G_{tot}(0)$ from calibration, we thus obtain $1/L_{tot}(\Phi)$ and $G_{tot}(\Phi)$ which are plotted in Fig. \ref{fig:invLr-Gr_compare}(a). At low temperatures, $1/L_{tot}$ is reduced and $G_{tot}$ is enhanced at phase $\pi$, whereas such effect is weakened at high temperatures, consistent with earlier results \cite{ziwei_gSNS,schoenenberger_gSNS,AuPd_SNS_admittance}. 

\noindent For the noise measurement, at each temperature, $S_V(f)$ is taken at a series of phases from 0 to $2\pi$. Fig. \ref{fig:invLr-Gr_compare}(b) shows two examples at phase 0 and $\pi$ for $T$ = 60 mK. The voltage noise on resonance decreases significantly at phase $\pi$, consistent with the enhanced $G_{tot}$. Using the calibrated $S_{I,amp}$ and the transmission measurement in Fig. \ref{fig:invLr-Gr_compare}(a) at each temperature, $S_{I,th}(f)$ is extracted according to Eq. \ref{eq:Sv-f}. As plotted in Fig. \ref{fig:invLr-Gr_compare}(c), $S_{I,th}(f)$ for a given phase is frequency-independent and shows the different mean value marked by the dashed line. To further reduce the uncertainty of such mean value, around 100 data points shown in the figure are averaged to calculate $S_{I,th}$ at the given phase. Such mean value (squares) with the associated standard error of mean (error bar) for all phases and temperatures are summarized in Figs. \ref{fig:invLr-Gr_compare}(d,e). Meanwhile, $4k_BTG_{tot}$ using the transmission data is plotted as the solid lines. For the whole data set, quantitative agreement between $S_{I,th}$ and $4k_BTG_{tot}$ is achieved. Its phase-dependent part is thus a direct demonstration of the FDT for the SNS ring alone. The very high precision of the measurement can be appreciated by the fact that we are able to resolve a current noise variation of 20 fA\textsuperscript{2}/Hz as well as a temperature difference of a few milli-Kelvin. The simultaneous satisfaction of the FDT for both the resonator and the SNS ring using the same temperature also indicates that the ABSs of the SNS ring are well thermalized with the electronic temperature of the resonator, which is otherwise hard to be confirmed experimentally.

\section*{Understanding temperature dependence of dissipation}

\begin{figure}[]
  \centering
    \includegraphics[width= 0.5\textwidth]{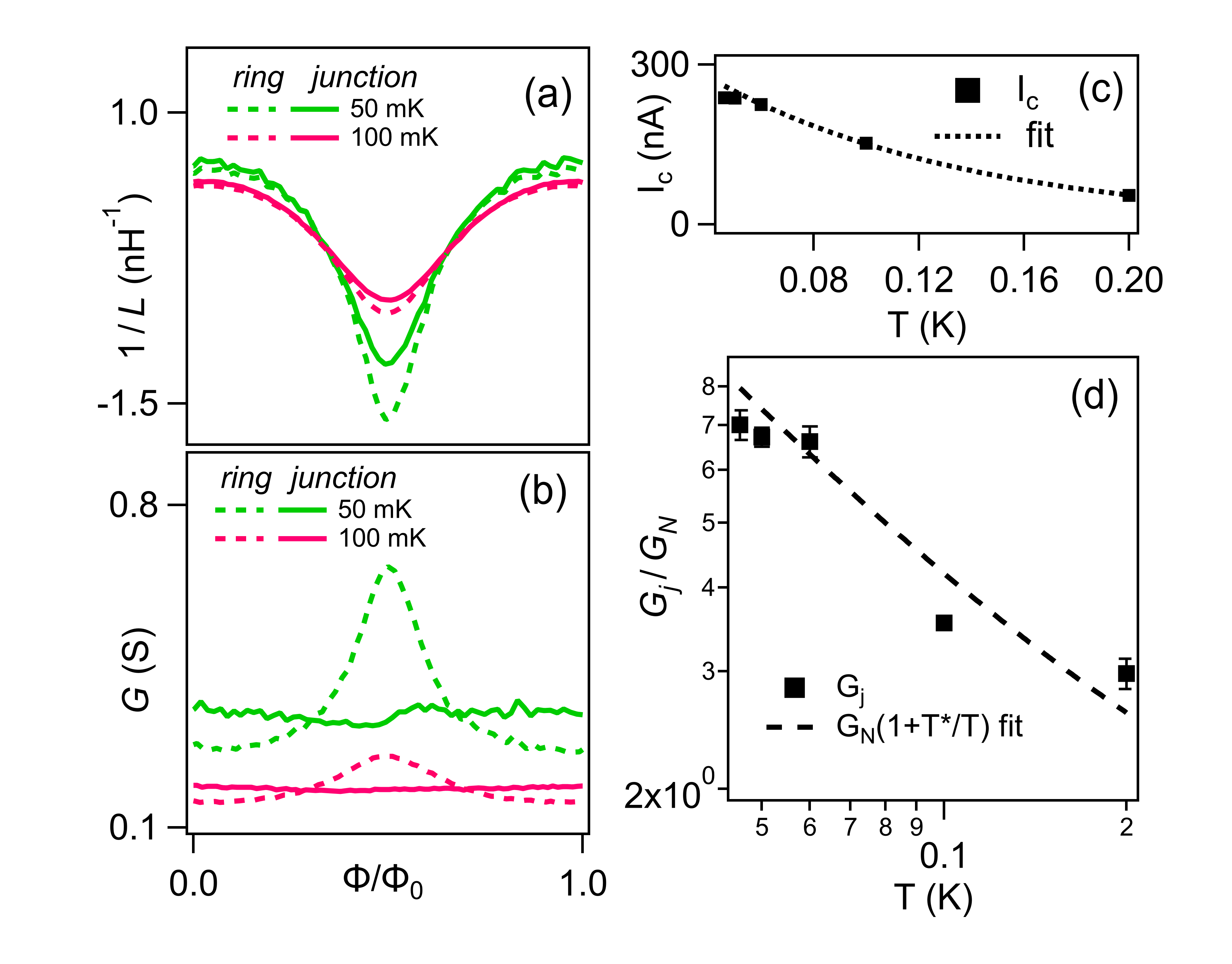}
\caption{Screening effect of SNS ring and temperature dependence of junction conductance: (a) $1/L_{ring}$ (dashed lines) and $1/L_j$ (continuous lines) after removing the screening effect. (b) Similar comparison for $G_{ring}$ (dashed) and $G_j$ (continuous). Only $T$ = 50 mK and 100 mK are shown for clarity. The data for other $T$ are shown in \cite{supp}. (c) Critical current $I_c$ versus $T$ (squares) is fitted to the long diffusive junction model \cite{dubos_Ic-T} (dashed line), giving $E_{TH}/k_B =$ 30 mK. (d) Normalized $G_j$ versus $T$ (squares). $G_j(T)$ is fitted to $G_N(1+T^*/T)$ (dashed line) with $G_N = $ 54 mS and $T^* =$ 30 mK.}
\label{fig:screen_chi}
\end{figure}

\noindent Having established the equivalence between the supercurrent noise and the conductance of the SNS ring, we focus on the conductance and try to understand its implied mechanism. The admittance of the SNS ring $Y_{ring}$ is reproduced without the resonator contribution in Figs. \ref{fig:screen_chi}(a, b) as the dashed lines. However, due to the screening effect of the loop supercurrent, $Y_{ring}$ is in general not equal to $Y_j$ \cite{bastien_prb} of an isolate junction. In order to facilitate comparison with theory developed for the junction, we thus convert from $Y_{ring}$ to $Y_j$ by $1/L_{ring} = (1/L_{j})/(1+L_l\chi_{j}')$ and $G_{ring} = G_{j}/(1+L_l\chi_{j}')^2$, where $\chi'_j = -1/L_j$ is the real component of the junction magnetic susceptibility and $L_l = 250$ pH is the loop inductance \cite{bastien_prb, supp}. Therefore, $1/L_{ring}$ (or $G_{ring}$) and $1/L_j$ (or $G_{j}$) are identical only when the screening coefficient $\beta(T) = L_l\chi_{j}'\ll 1$. In our case, $\beta \sim$ 0.1 at low $T$, and the screening effect is not negligible. The junction $1/L_j$ and $G_j$ converted from the ring $1/L_{ring}$ and $G_{ring}$ are plotted in Figs. \ref{fig:screen_chi}(a, b) as the solid lines. For the inverse inductance, the correction at low temperature is sizeable while at high temperature it is negligible due to the reduced supercurrent. By integrating $1/L_j = \partial I/\partial \Phi$, the critical current $I_c$ is obtained and plotted in Fig. \ref{fig:screen_chi}(c). From $I_c(T)$, we estimate the Thouless energy $E_{TH}$ to be 30 mK \cite{dubos_Ic-T, supp}, and the temperature in the experiment is thus always higher than $E_{TH}$. For conductance, on the other hand, $G_j$ is almost phase-independent for all temperatures. We emphasize that such phase-independent $G_j$ cannot be confused with $G_{reso}$ since $G_{reso}$ alone does not produce the observed phase dependence in $\delta (1/Q)$\cite{supp, bastien_prb}. Instead, it is the screening effect that enables us to determine the flux-independent $G_j$ which is otherwise difficult to be separated from $G_{reso}$ obtained in Fig. \ref{fig:invLr-Gr_compare}(a) \cite{supp, bastien_prb}. In Fig. \ref{fig:screen_chi}(d), $G_j(T)$ shows strong enhancement at low temperature. This behavior is drastically different from the Drude conductance of an incoherent normal metal, which is temperature-independent in the milli-Kelvin range. $G_j(T)$ can also be fitted to the relation $G_N(1+T^*/T)$ according to other works \cite{bastien_prb,tikhonov_admittance,spivakdissipation}, giving $G_N$ = 54 mS close to the Drude conductance of a similar control junction and $T^*$ = 30 mK close to the Thouless energy $E_{TH}$ estimated from $I_c(T)$.

\begin{figure}[]
  \centering
    \includegraphics[width= 0.5\textwidth]{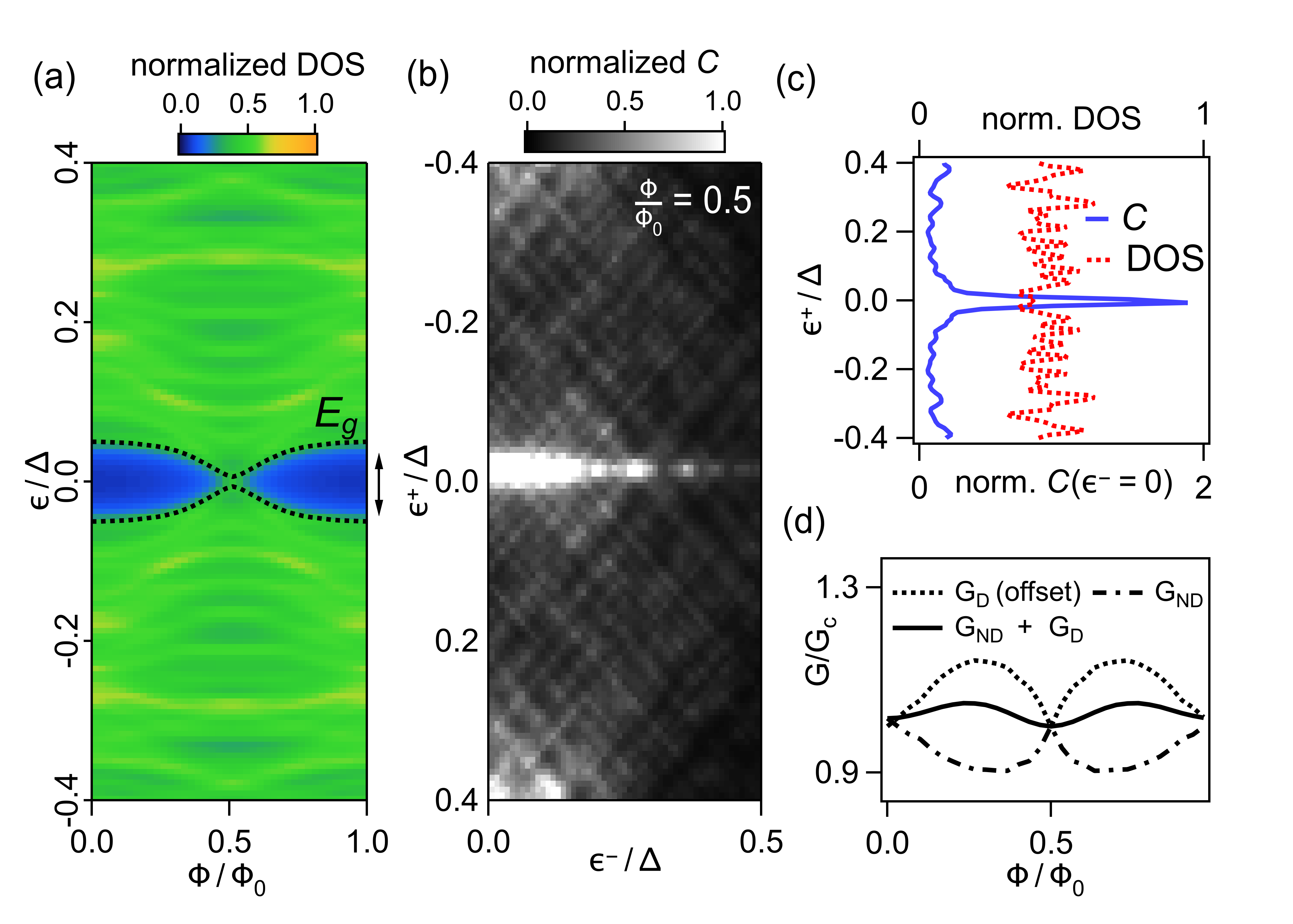}
\caption{Theory to explain $G_j(\Phi,T)$: (a) Density of state versus flux. $\Delta$ is the superconducting gap of the electrode. $E_g$ is the minigap. (b) Current correlation function $\mathcal{C}(\epsilon^+, \epsilon^-)$ at $\Phi/\Phi_0 = $ 0.5. (c) $\mathcal{C}(\epsilon^+, \epsilon^-=0)$ (blue solid) and DOS (red dashed) at $\Phi/\Phi_0 = 0.5$. The peak of $\mathcal{C}$ at $\epsilon^+ = 0$ is essential to $G_j \sim 1/T$ (see text). (d) The total conductance $G_j$ (solid) and its two components $G_D$ (dashed) and $G_{ND}$  (dot-dashed), normalized by $G_c = G(0.5\Phi_0)$. $G_D$ is offset by $G_c$. $h\gamma = E_g$ and $k_BT = 4E_g$. The aspect ratio $L/W$ = 8; the superconducting coherence length and the mean-free path $\xi, l_e \approx L/10$. $\Delta/E_g \approx 10$. See \cite{supp}.}
\label{fig:simu}
\end{figure}

\noindent To explain the weak phase dependence as well as the strong temperature dependence of $G_j$, we recall that by the linear response theory $G_j = G_D + G_{ND}$ can be written in two terms \cite{bastien_prb,trivedi-browne_chi,buettikernoise}. In \cite{supp} we justify that the resonance frequency is much less than the inelastic scattering rate $\gamma$ and $k_BT/h$. Therefore, taking the zero-frequency limit, we have:

\begin{align}
\begin{split}
G_D &= -\frac{1}{h\gamma}\sum_n |J_{nn}|^2\frac{\partial f_n}{\partial E_n}\\
G_{ND} &= -\sum_{n\neq m} |J_{nm}|^2\frac{h\gamma}{(E_n-E_m)^2+(h\gamma)^2}\frac{f_n-f_m}{E_n-E_m}\\
\end{split}
\label{eq:Kubo} 
\end{align}

\noindent where $E_n(\Phi)$ is the n-th Andreev level, $f_n = f(E_n)$ is the Fermi-Dirac function and $\gamma$ is the inelastic scattering rate. $J_{nm} = (ie\hbar/m^*)\langle n| \nabla|m \rangle$ is the matrix element of the current operator. The term $G_D$ only involves the diagonal element of $J$ while the term $G_{ND}$ involves all the non-diagonal elements. $E_n$ and $J_{nm}$ for long-diffusive junction can be numerically calculated by diagonalizing the Bogoliubov-de Gennes Hamiltonian of the junction \cite{meydi_chi_theory,supp}. The density of state (DOS) versus $\Phi$ calculated from $E_n(\Phi)$ is plotted in Fig. \ref{fig:simu}(a), showing the phase-dependent minigap $E_g$ expected for a long-diffusive junction (dashed line) \cite{ZhouSpivak_minigap}. We first explain the weak phase dependence of $G_j$. In the experiment where $k_BT \gg E_{g}$, many ABSs transitions across $E_g$ occur due to thermal excitation. Therefore, it is justified to assume that all $|J_{nm}|^2$ with significant values are included in $G_{ND}$ and $G_j \propto \sum(|J_{nn}|^2+|J_{nm}|^2) = Tr(|J|^2)$ which does not depend on the Aharanov-Bohm phase \cite{bastien_prb}. Indeed, the phase dependences of $G_D$ and $G_{ND}$ cancel each other, resulting in a weakly phase-dependent $G_j$ reproduced by calculation in Fig. \ref{fig:simu}(d). We also note that $h\gamma$ cannot be much smaller than $E_g$ to have a phase-independent $G_j$ \cite{supp}. With $E_g/k_B$ in the order of 10 mK, $\gamma$ is in the order of 100 MHz, reasonable for similar set-ups \cite{bastien_prb,anil_microwave}.

\noindent To explain the strong temperature dependence of $G_j$, we first approximate $G_{ND}$ in the continuous spectrum limit, since the level spacing of the ABSs in our junction is the smallest energy scale \cite{supp}:


\begin{equation}
\begin{split}
   &G_{ND} \propto \iint \,d\epsilon^+\,d\epsilon^-\,\mathcal{C}(\epsilon^+,\epsilon^-)\mathcal{F}(\epsilon^+,\epsilon^-)\\
   &\mathcal{F}(\epsilon^+,\epsilon^-) = \frac{h\gamma}{(\epsilon^-)^2+(h\gamma)^2}\frac{f(\epsilon^++\frac{\epsilon^-}{2})-f(\epsilon^+-\frac{\epsilon^-}{2})}{\epsilon^-}\\
\end{split}
\label{eq:normalGnd} 
\end{equation}

\noindent where $\epsilon^+$ and $\epsilon^-$ are the average and the difference between two energy levels, respectively, and $\mathcal{C}(\epsilon^+,\epsilon^-)$ is the current correlation function convoluting the DOS and the matrix elements of the current operator $|J_{nm}|^2$ (for the full definition see \cite{supp}). It is illustrative to briefly examine the case of an unproximitized normal metal. Here $G_D \equiv 0$ since $E_n$ is phase-independent and $J_{nn} = \partial E_n/\partial \varphi \equiv 0$. For $G_{ND}$, since $\mathcal{C}(\epsilon^+,\epsilon^-)$ is constant in energy \cite{supp}, the integration involving the Fermi-Dirac distribution in Eq. \ref{eq:normalGnd} thus gives a temperature-independent conductance. The result can be linked to the Drude conductance if $\mathcal{C}$ includes the elastic scattering \cite{supp}. The situation is drastically different for the proximitized normal metal. From Eq. \ref{eq:normalGnd}, the key condition for a temperature-dependent $G_{ND}$ is an energy-dependent $\mathcal{C}(\epsilon^+,\epsilon^-)$. As discussed in previous works \cite{spivakdissipation,Pauli_ac_chi}, the presence of the minigap at phase 0 provides such energy-dependence of $\mathcal{C}$, resulting $G_j \sim 1/T$. In our case, since $G_j$ is phase-independent, it is thus surprising to see that such $1/T$ dependence is also true even at phase $\pi$ where the minigap closes. To explain this, the maps of $\mathcal{C}(\epsilon^+,\epsilon^-)$ at phase $\pi$ is shown in Fig. \ref{fig:simu}(b), with line cuts along $\epsilon^- = 0$ shown in Fig. \ref{fig:simu}(c) (blue solid). Although the DOS is indeed almost constant (red dashed), a sharp peak in $\mathcal{C}$ appears at $\epsilon^+ = 0$ for a wide range of $\epsilon^-$, revealing a strong current correlation between the ABSs symmetrical across the Fermi level, even for levels far in energy \cite{meydi_chi_theory}. In \cite{supp}, we show that such feature does not exist for a normal metal without superconducting contacts to impose the electron-hole symmetry for energy smaller than $\Delta$. Since the peak width is much smaller than $h\gamma$ and $k_BT$, it can be approximated as $\delta(\epsilon^+)$. Inserting in Eq. \ref{eq:normalGnd} and using $G_D = 0$ at phase $\pi$, we thus arrive at $G_j = G_{ND} \propto \int \,d\epsilon^-\,\mathcal{F}(0,\epsilon^-)$. In the experiment where $k_BT\gg h\gamma$, the expression can be further approximated by $G_j \sim \partial f/\partial \epsilon'(\epsilon'=0) \sim 1/T$ thus giving the observed temperature dependence. This shows that the electron-hole symmetry imposed by the superconducting contacts makes a proximitized normal metal fundamentally different from an unproximitized one even for phase $\pi$ where the minigap closes.   

\noindent In conclusion, we have measured the supercurrent noise as well as the linear admittance of a phase-biased SNS ring using a superconducting resonator. We verify the FDT long-predicted for the SNS junction for all phases and temperatures. In addition, the junction conductance $G_j$ is several times larger than the Drude conductance and shows an $1/T$ dependence for all phases, in particular at phase $\pi$. This highlights a key difference between a proximitized and unproximitized metal, even when the minigap closes. Using the linear response theory, we argue that $G_j \sim 1/T$ is a manifestation of the enhanced current correlation between the ABSs symmetrical across the Fermi level due to the superconducting proximity effect. Our results provides insight to the nature of the noise in the hybrid superconducting systems, many of which are promising candidates for ultrasensitive quantum devices. They may also be applied to the more exotic systems such as the topological SNS junction, whose supercurrent noise can reveal the evidence of a topologically protected crossing between the Andreev levels \cite{fu-kane-noise}. 




%

\end{document}